\documentclass[aps,showpacs,superscriptaddress]{revtex4}
\usepackage{graphicx}
\usepackage{amssymb,amsmath}
\begin{document}

\title{\bf\boldmath Study of the process $e^+e^-\to\omega\eta\pi^0$ in the energy range
$\sqrt{s} <2$ GeV with the SND detector}

\author{M.~N.~Achasov}
\author{V.~M.~Aulchenko}
\author{A.~Yu.~Barnyakov}
\author{K.~I.~Beloborodov}
\author{A.~V.~Berdyugin}
\email[e-mail:]{berdugin@inp.nsk.su}
\author{D.~E.~Berkaev}
\affiliation{Budker Institute of Nuclear Physics, SB RAS, Novosibirsk, 630090,
Russia}
\affiliation{Novosibirsk State University, Novosibirsk, 630090, Russia}
\author{A.~G.~Bogdanchikov}
\author{A.~A.~Botov}
\affiliation{Budker Institute of Nuclear Physics, SB RAS, Novosibirsk, 630090,
Russia}
\author{T.~V.~Dimova}
\author{V.~P.~Druzhinin}
\author{V.~B.~Golubev}
\author{L.~V.~Kardapoltsev}
\author{A.~G.~Kharlamov}
\affiliation{Budker Institute of Nuclear Physics, SB RAS, Novosibirsk, 630090,
Russia}
\affiliation{Novosibirsk State University, Novosibirsk, 630090, Russia}
\author{I.~A.~Koop}
\affiliation{Budker Institute of Nuclear Physics, SB RAS, Novosibirsk, 630090,
Russia}
\affiliation{Novosibirsk State University, Novosibirsk, 630090, Russia}
\affiliation{Novosibirsk State Technical University, Novosibirsk, 630092, Russia}
\author{A.~A.~Korol}
\author{D.~P.~Kovrizhin}
\affiliation{Budker Institute of Nuclear Physics, SB RAS, Novosibirsk, 630090,
Russia}
\affiliation{Novosibirsk State University, Novosibirsk, 630090, Russia}
\author{S.~V.~Koshuba}
\author{A.~S.~Kupich}
\author{A.~P.~Lysenko}
\author{N.~A.~Melnikova}
\author{K.~A.~Martin}
\author{E.~V.~Pakhtusova}
\author{A.~E.~Obrazovsky}
\affiliation{Budker Institute of Nuclear Physics, SB RAS, Novosibirsk, 630090,
Russia}
\author{E.~A.~Perevedentsev}
\author{Yu.~A.~Rogovsky}
\author{S.~I.~Serednyakov}
\author{Z.~K.~Silagadze}
\author{Yu.~M.~Shatunov}
\author{P.~Yu.~Shatunov}
\author{D.~A.~Shtol}
\affiliation{Budker Institute of Nuclear Physics, SB RAS, Novosibirsk, 630090,
Russia}
\affiliation{Novosibirsk State University, Novosibirsk, 630090, Russia}
\author{A.~N.~Skrinsky}
\affiliation{Budker Institute of Nuclear Physics, SB RAS, Novosibirsk, 630090,
Russia}
\author{I.~K.~Surin}
\author{Yu.~A.~Tikhonov}
\author{ Yu.~V.~Usov}
\author{A.~V.~Vasiljev}
\author{I~M.~Zemlyansky}
\affiliation{Budker Institute of Nuclear Physics, SB RAS, Novosibirsk, 630090,
Russia}
\affiliation{Novosibirsk State University, Novosibirsk, 630090, Russia}

\begin{abstract}
The process $e^+e^-\to\omega\eta\pi^0$ is studied in the
energy range $1.45-2.00$ GeV using data with an integrated
luminosity of 33 pb$^{-1}$ accumulated by the SND 
detector at the $e^+e^-$ collider VEPP-2000.  The $e^+e^-\to\omega\eta\pi^0$
cross section is measured for the first time. The cross section has a 
threshold near 1.75 GeV. Its value is about 2 nb in the energy range 
$1.8-2.0$ GeV. The dominant intermediate state for the process
$e^+e^- \to \omega\eta\pi^0$ is found to be $\omega a_0(980)$.
\end{abstract}
\pacs{  
13.66.Bc            
14.40.Be            
13.40.Gp            
12.40.Vv}           
\maketitle

\section{Introduction}
This work continues the study of multiphoton processes $e^+e^- \to n\gamma$ in the 
center-of-mass (c.m.) energy domain $\sqrt{s}<2$ GeV with the SND 
detector~\cite{ompi,etag,etap,eta}. The main goal of these studies is the
measurement of radiative decays of excited vector resonances of the 
$\rho$, $\omega$ and  $\phi$ families~\cite{etag}, as well as the search for 
rare processes of $C$-even resonance production in the $e^+e^-$ 
annihilation~\cite{etap,eta}. While searching for the rare reactions mentioned 
above, hadronic processes containing $\omega$-meson in the final state 
decaying into  $\pi^0\gamma$ constitute a significant background (the 
$\omega\to\pi^0\gamma$ branching fraction is $(8.28\pm0.28)\%$~\cite{pdg}).
For example, the process
$e^+e^-\to \omega\pi^0\to\pi^0\pi^0\gamma$~\cite{ompi} dominates in the 
five-photon final state and hinders the search for the radiative processes
$e^+e^-\to f_0\gamma$, $f_2\gamma$.

The $e^+e^- \to \eta\pi^0\pi^0\gamma$ process studied in this work is 
important to search for the radiative processes $e^+e^- \to \eta^\prime
\gamma$ and $f_1(1285)\gamma$. A possible source of hadronic background in the 
$\eta\pi^0\pi^0\gamma$ final state is the process $e^+e^- \to \omega\eta\pi^0$.

At energies below 2 GeV, the total cross section of $e^+e^-$ annihilation into 
hadrons needed, for example, to calculate the running coupling constant of 
the electromagnetic interactions, is determined as a sum of exclusive 
hadronic cross sections. The process $e^+e^- \to \omega\eta\pi^0$ has not 
previously been measured and was not included in this sum.
In this work we select this process for the first time and
measure its cross section.

\section{Detector and experiment}
The SND detector collects data at the $e^+e^-$ collider
VEPP-2000~\cite{VEPP2000} operating at c.m. energies 
$\sqrt{s}=0.3-2.0$ GeV.  This analysis uses data collected in 
2010-2012. During the experiments, the energy range 1.05-2.00 GeV was scanned
several times with a step of 20-25 MeV. Because of the smallness of 
statistics, we measure the cross section averaged over the energy intervals, 
listed in Table~\ref{tabl1}. 

A detailed description of the SND detector is given in 
Refs.~\cite{snd1,snd2,snd3,snd4}. The main part of this non-magnetic detector 
is a three-layer spherical electromagnetic calorimeter based on NaI(Tl) 
crystals. The solid angle coverage of the calorimeter is 95\% of 4$\pi$. 
Its energy resolution for photons is 
$\sigma_{E_\gamma}/E_{\gamma}=4.2\%/\sqrt[4]{E_\gamma({\rm GeV})}$, 
and the angular resolution is about $1.5^\circ$. 
Direction of charged particles are measured in a tracking system
consisting of a nine-layer drift chamber and a proportional chamber with
the signal readout from the cathode strips. The solid angle coverage of 
the tracking system is 94\% of 4$\pi$.  From the outside the SND calorimeter 
is surrounded by a muon system. In this analysis, the muon system veto is used 
to suppress the cosmic-ray background.

Simulation of the signal and background processes is performed 
using Monte-Carlo generators that take into account initial-state radiative 
corrections calculated according to Ref.~\cite{radcor}. In particular, emission
of an additional photon is simulated with the angular distribution according 
to Ref~\cite{BM}. Interactions of the particles produced in the 
$e^+e^-$ annihilation with the detector material are modeled using the GEANT4 
package~\cite{GEANT4}. 
Simulation takes into account changes in the experimental conditions during 
the data taking, in particular, dead detector channels 
and variations of the beam-induced background. The beam 
background leads to the appearance of spurious photons and charged tracks in 
data events. To account for this effect in the simulation, 
special background events, recorded during the experiment with a random 
trigger, are used. Fired detector channels in these events are 
superimposed on the simulated events.

In this work, the $e^+e^-\to\omega\eta\pi^0$ process is studied in the channel 
$e^+e^-\to \eta\pi^0\pi^0\gamma\to 7\gamma$. Since the final state for the 
process under study contains no charged particles, we use the process  
$e^+e^-\to \gamma\gamma$ for normalization. As a result of the normalization 
a part of systematic uncertainties associated with the hardware event selection
and spurious charged tracks from the beam background are canceled out. 
Accuracy of the luminosity 
measurement using the $e^+e^-\to \gamma\gamma$ process is 2.2\%~\cite{ompi}.

\section{Selection criteria}
The selection of signal $e^+e^-\to \eta\pi^0\pi^0\gamma\to 7\gamma$ events is
performed in two stages. Initially, events with exactly 7 photons with energy
greater than 20 MeV, no charged particles, and for which the muon-system
veto is not triggered, are selected. For these events,
the following conditions on the total energy deposition in the calorimeter 
$E_{\rm EMC}$ and on the total event momentum $P_{\rm EMC}$, calculated using 
energy depositions in calorimeter crystals, are imposed:   
\begin{equation}
\label{sel1}
0.7 < E_{\rm EMC}/\sqrt{s} < 1.2,~ P_{\rm EMC}/\sqrt{s} < 0.3,~
(E_{\rm EMC} - P_{\rm EMC})/\sqrt{s} > 0.7.
\end{equation}
The transverse profile of the energy deposition in the calorimeter for 
reconstructed photons is required to be consistent with  that expected for an 
electromagnetic shower~\cite{xinm}. The latter requirement provides separation
of events with well isolated photons from those
with merged photons or with clusters in the calorimeter produced by $K_L$
mesons.

The main background processes are the following: 
$e^+e^-\to \omega\pi^0 \to \pi^0\pi^0\gamma$,
$e^+e^-\to\omega\pi^0\pi^0,\eta\gamma \to \pi^0\pi^0\pi^0\gamma$, 
$e^+e^-\to\omega\eta\to\eta\pi^0\gamma$ and 
$e^+e^-\to K_SK_L\pi^0$ with the decay $K_S\to\pi^0\pi^0$. 
The process $e^+e^-\to \pi^0\pi^0\gamma$, having five photons in the final 
state, is a background source because of a relatively large cross
section. Additional photons in $e^+e^-\to \pi^0\pi^0\gamma$ 
events arise from splitting of electromagnetic showers,  
initial state radiation, and beam-induced background.

\begin{figure}
\includegraphics[width=0.4\textwidth]{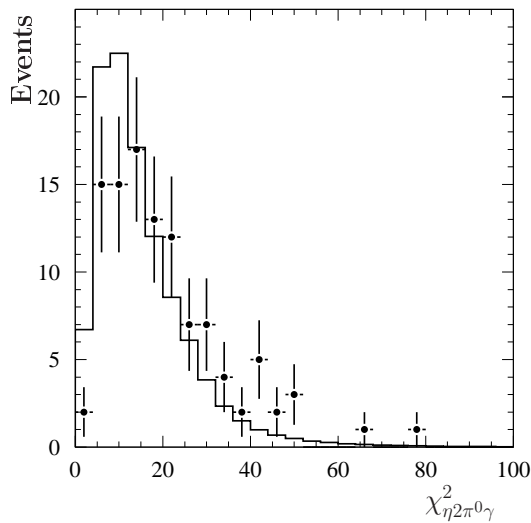}
\caption{ \label{fig:x2et2p0g_wo2pg_wo3pg}
The $\chi^2_{\eta 2\pi^0\gamma}$ distributions for selected data
events (points with error bars) and for simulated 
$e^+e^-\to\eta 2\pi^0\gamma$ events (histogram).}
\end{figure}

\begin{figure}
\includegraphics[width=0.4\textwidth]{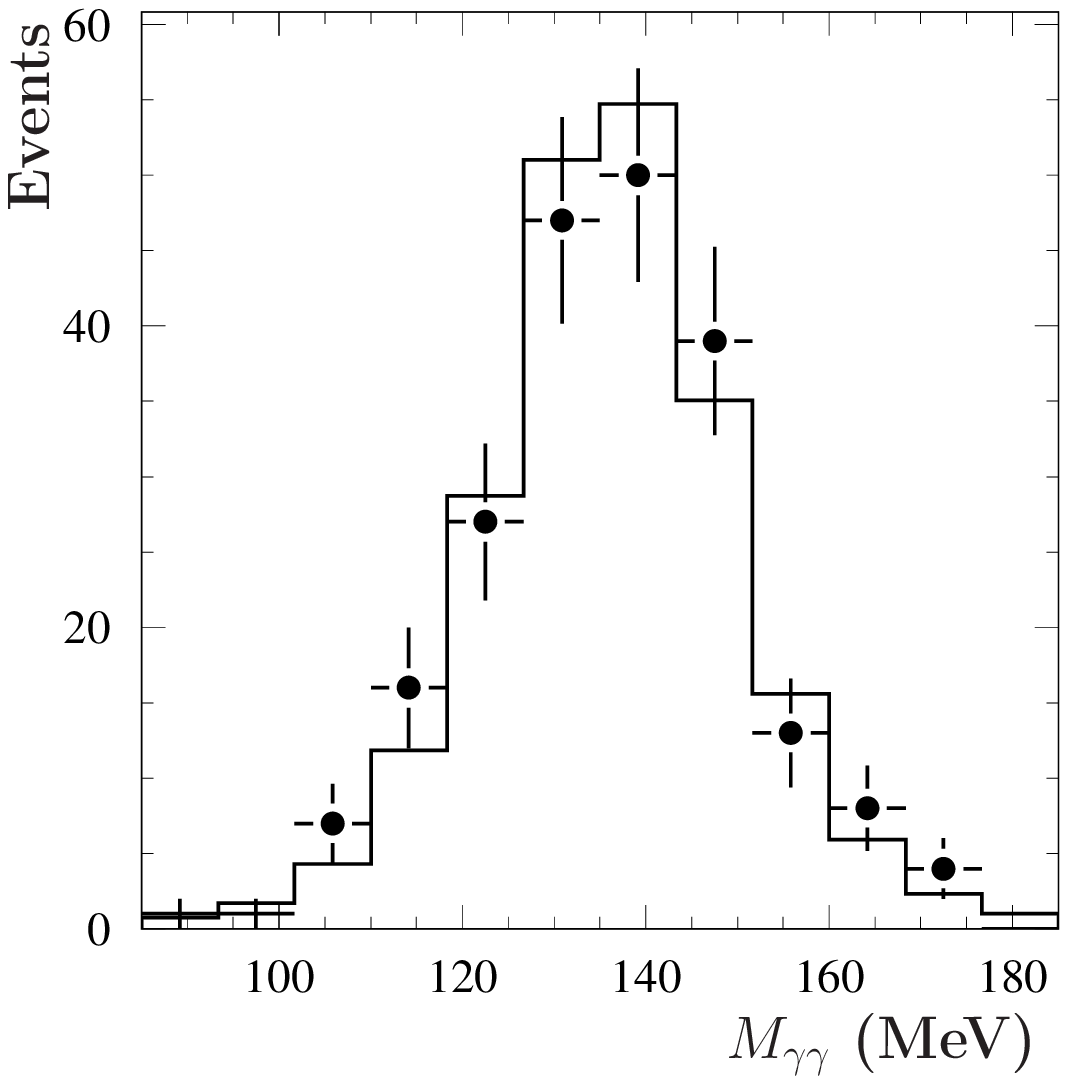}
\includegraphics[width=0.4\textwidth]{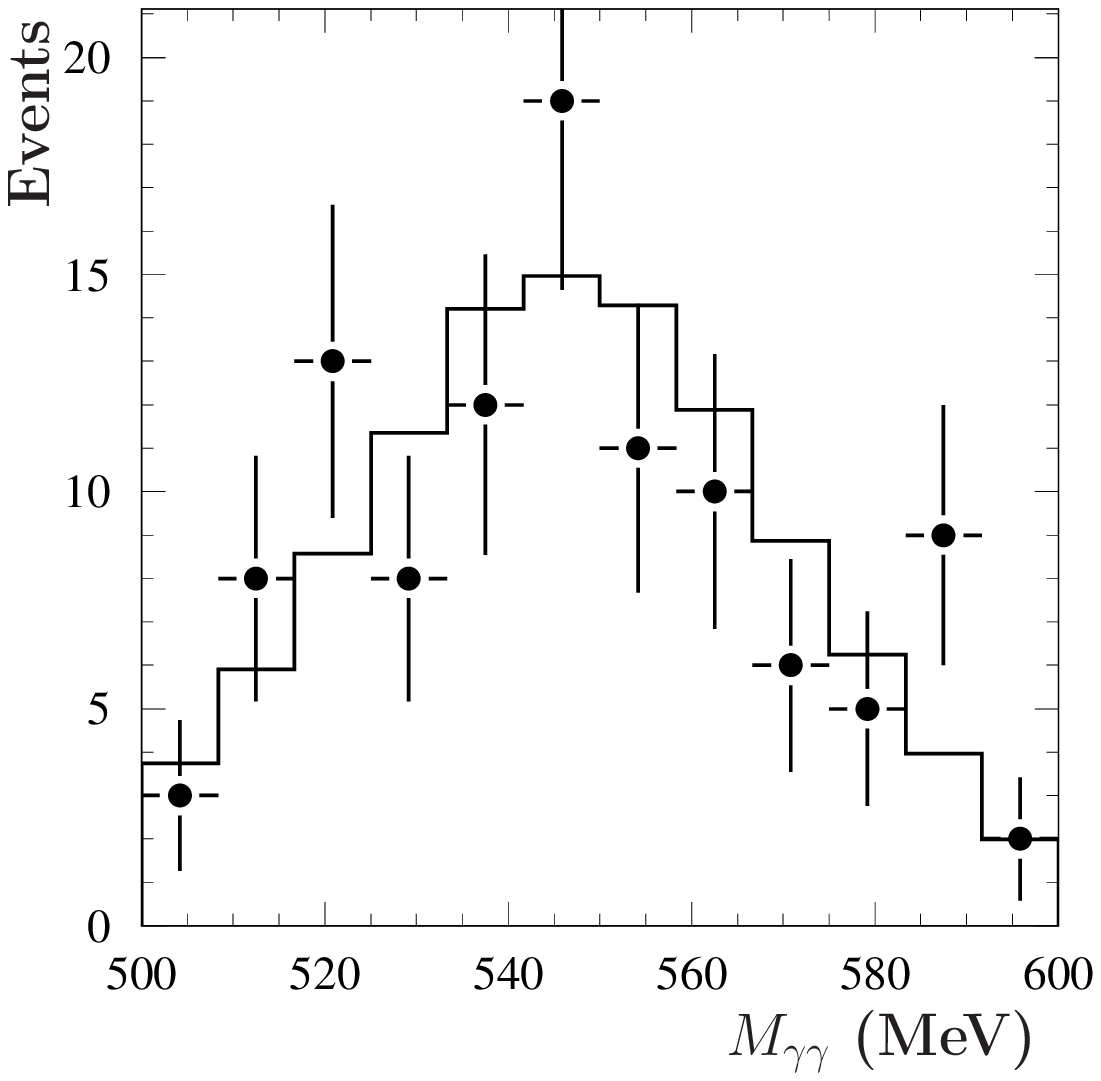}
\caption{ \label{fig:mgg_7g_wo2pg_wo3pg}
The distributions of the two-photon invariant mass of $\pi^0$-meson candidates
(left, two entries per events) and of $\eta$-meson candidates (right) for 
selected data events (points with error bars) and for simulated 
$e^+e^-\to\eta 2\pi^0\gamma$ events
(histogram).}
\end{figure}

Further selection of events is based on kinematic fits,
which use the measured photon angles and energies as input parameters.
The fit is performed under the hypothesis that the seven-photon event proceed
through the particular set of intermediate particles (with corresponding
mass constraints), and satisfies 
energy-momentum conservation laws. As a result of the kinematic fit,
photon energies are refined and the $\chi^2$ of the assumed kinematic 
hypothesis is calculated. First of all, the $e^+e^-\to 7\gamma$ hypothesis
is tested, and the condition $\chi^2_{7\gamma}<30$ is imposed. Then the photon
pairs, candidates for the $\pi^0$ and $\eta$ mesons, are searched. It is 
required that the invariant mass of the candidate is in the 
range $(m_{\pi^0,\eta}-50\mbox{ MeV},m_{\pi^0,\eta}+50\mbox{ MeV})$.
Events with one $\eta$ meson candidate and two $\pi^0$ candidates are
selected as possible signal events. To suppress the 
background from the processes $e^+e^-\to\omega\pi^0\pi^0,\eta\gamma \to 
\pi^0\pi^0\pi^0\gamma$, events containing three $\pi^0$ candidates 
are rejected. A kinematic fit is also performed to the 
$e^+e^-\to\pi^0\pi^0\gamma$ hypothesis.  All five-photon combinations with 
two $\pi^0$-meson candidates are tested. Events with 
$\chi^2_{\pi^0\pi^0\gamma} < 50$ are rejected. For remaining events, 
a kinematic fit is performed to the $e^+e^-\to\eta 2\pi^0\gamma$ hypothesis. 
The $\chi^2_{\eta 2\pi^0\gamma}$ distributions for data and 
simulated $e^+e^-\to\eta 2\pi^0\gamma$ events are shown in
Figure~\ref{fig:x2et2p0g_wo2pg_wo3pg}.
Figure~\ref{fig:mgg_7g_wo2pg_wo3pg} shows the distributions of the two-photon
invariant mass for $\pi^0$ and $\eta$ meson candidates for selected 
data and simulated $e^+e^-\to\eta 2\pi^0\gamma$ events.
It is evident from the distributions in Figs.~\ref{fig:x2et2p0g_wo2pg_wo3pg} and
\ref{fig:mgg_7g_wo2pg_wo3pg} that most selected events arise from 
the process $e^+e^-\to\eta 2\pi^0\gamma$.

\begin{figure}
\includegraphics[width=0.4\textwidth]{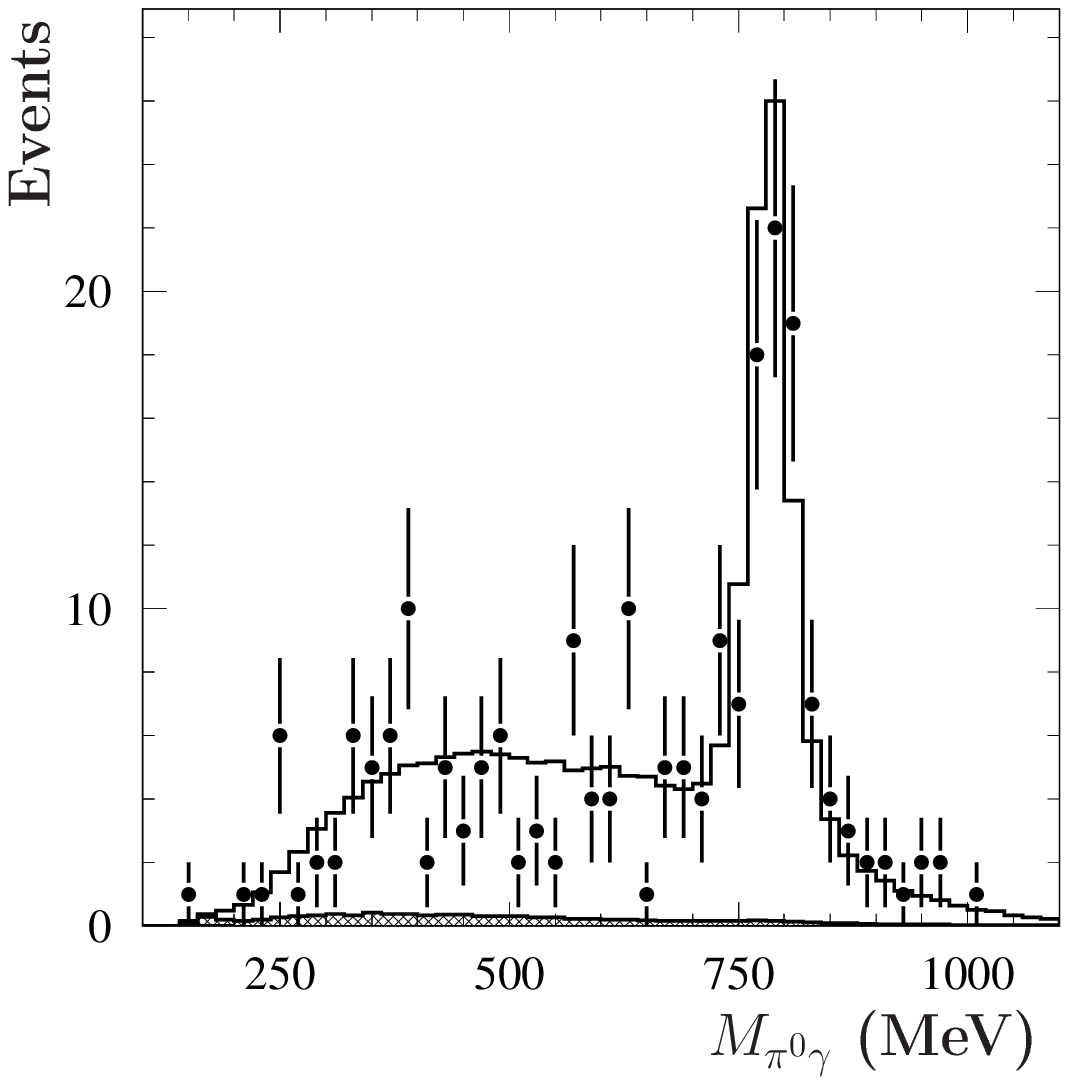}
\caption{ \label{fig:mpg_7g_wo2pg_wo3pg_wet2p0g}
The $M_{\pi^0\gamma}$ spectrum for selected $e^+e^-\to \eta 2\pi^0\gamma$ 
candidate events (points with error bars, two entries per event). 
The histogram is the sum  
of the distributions for simulated $e^+e^-\to\omega\eta\pi^0$ and  
background events. 
The simulated distribution is normalized to the number of data events. The 
shaded histogram shows the background distribution.}
\end{figure}

Figure~\ref{fig:mpg_7g_wo2pg_wo3pg_wet2p0g} shows the distribution of
the $\pi^0\gamma$ invariant mass ($M_{\pi^0\gamma}$) for
106  $e^+e^-\to \eta 2\pi^0\gamma$ candidate events.
The calculated background from the processes listed above is 9.2 events
(3 events from $\omega\eta$, 3 events from $\omega\pi^0\pi^0$,
2 events from $K_SK_L\pi^0$, 1 event from $\eta\gamma$). To calculate 
the cross sections of these background processes,
we use the results of Refs.~\cite{BABAR6pi,BABARomegapipi,BABARKKpi,etag} and 
isotopic relations.

The histogram in Fig.~\ref{fig:mpg_7g_wo2pg_wo3pg_wet2p0g} shows the 
simulated distribution for the sum of the signal and estimated
background under the assumption that the signal is from the 
process $e^+e^-\to\omega\eta\pi^0$. The simulated distribution is normalized to
the number of selected data events. It is seen that at the existing 
statistical level these two contributions are sufficient to describe the
distribution of selected $e^+e^-\to \eta 2\pi^0\gamma$ candidate events.

For the final selection of $e^+e^-\to\omega\eta\pi^0$ events,
the condition $|M_{\pi^0\gamma}-M_{\omega}|<50$~MeV is required  
for at least one $\pi^0\gamma$ combination in an event. This condition is 
satisfied by 62 events. Their distribution over the energy intervals is given 
in Table~\ref{tabl1}. The estimated number of background events is equal to 
0.9. The systematic uncertainty of background calculation is 
taken to be 100\%.

\begin{figure}
\includegraphics[width=0.4\textwidth]{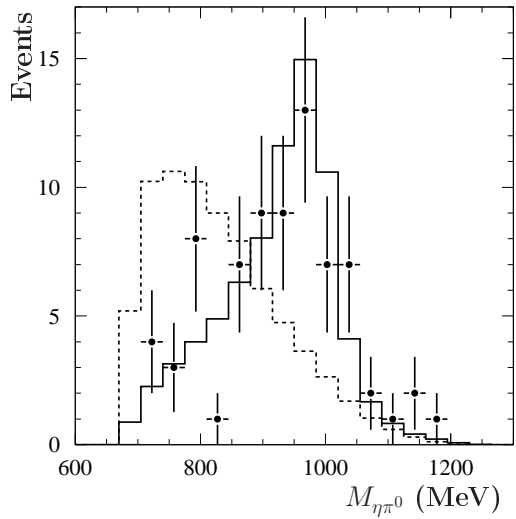}
\caption{ \label{fig:metp0_fullsel}
The $M_{\eta\pi^0}$ spectrum for selected $e^+e^-\to\omega\eta\pi^0$ candidate 
events (points with error bars). The solid histogram represents  
$e^+e^-\to\omega a_0(980)$ simulation, while the dashed histogram represents
simulation of the process $e^+e^-\to\omega\eta\pi^0$ with uniform phase-space 
distribution of the final particles.}
\end{figure}

The spectrum of the $\eta\pi^0$ invariant mass ($M_{\eta\pi^0}$) for $e^+e^-
\to\omega\eta\pi^0$ candidate events is shown in Fig.~\ref{fig:metp0_fullsel}. 
The $\pi^0$ candidate with the maximal difference 
$|M_{\pi^0\gamma}-M_{\omega}|$ is used to calculate $M_{\eta\pi^0}$. 
For comparison, we also show the 
spectra for the simulated events of the process $e^+e^-\to\omega a_0(980)$
with the decay $a_0(980)\to\eta\pi^0$ and of the process
$e^+e^-\to\omega\eta\pi^0$ with uniform phase-space distribution of the final 
particles. It is evident that the data $M_{\eta\pi^0}$ spectrum
is consistent with the distribution for the $\omega a_0(980)$ model.

\section{Detection efficiency}
The detection efficiency for the events of the process 
$e^+e^-\to\omega a_0(980)\to\omega\eta\pi^0\to\eta\pi^0\pi^0\gamma\to 7\gamma$
is determined using MC simulation. The simulation takes into account the 
initial state radiative corrections~\cite{radcor}, in particular, the emission 
of additional photons.

\begin{figure}
\includegraphics[width=0.9\textwidth]{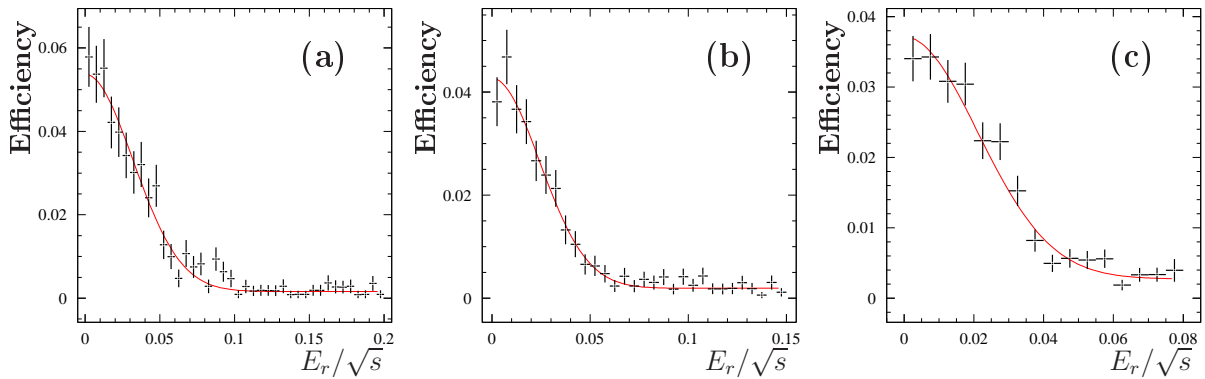}
\caption{ \label{fig:eff_ometp0_fullsel}
The dependence of the detection efficiency for
$e^+e^-\to \omega a_0(980)\to\omega\eta\pi^0\to\eta\pi^0\pi^0\gamma\to 7\gamma$
events on the energy of the additional photon emitted from the initial state
for $\sqrt{s} = 2.0$ GeV (a), 1.76 GeV (b) and  1.6 GeV (c). The points 
with error bars are obtained from simulation. The curve shows the result of 
approximation of the $\epsilon_r(\sqrt{s},E_r)$ dependence by a smooth 
function.}
\end{figure}

The detection efficiency $\epsilon_r$ is determined as a function of two
parameters: the c.m. energy $\sqrt{s}$ and the energy of the
additional photon $E_r$ emitted from the initial state. 
Figure~\ref{fig:eff_ometp0_fullsel} shows 
the dependence of the detection efficiency on $E_r$ for three 
representative c.m. energies.
The values of the detection efficiency at $E_r=0$ averaged over the energy 
intervals are given in Table~\ref{tabl1}.

To estimate the systematic uncertainty of the detection efficiency 
determination, we use the results of Ref.~\cite{etag} where
the difference in the detector responses between data and simulation 
for seven-photon events was studied. Based on this study, the systematic 
uncertainty on the detection efficiency is estimated to be 3\%.

\section{\boldmath Born cross section for the reaction $e^+e^-\to\omega\eta\pi^0$
\label{fit}}
The visible cross section for the process $e^+e^-\to\omega\eta\pi^0$ is 
related to the Born cross section ($\sigma(E)$) by the following formula:
\begin{equation}
\label{viscrs}
\sigma_{vis}(s) = B\int\limits_{0}^{x_{max}} 
\epsilon_r(\sqrt{s},\frac{x\sqrt{s}}{2}) F(s,x,) \sigma(s(1-x))dx,
\end{equation}
where $F(s,x)$ is the so called radiator function describing
the probability of radiating a certain energy fraction $x=2E_r/\sqrt{s}$ 
carried away by photons emitted
from the initial state~\cite{radcor}, and 
$B$ is a product of branching fractions 
$B={\cal B}(\omega\to\pi^0\gamma){\cal B}(\eta\to\gamma\gamma)
{\cal B}(\pi^0\to\gamma\gamma){\cal B}(\pi^0\to\gamma\gamma)$~\cite{pdg}.
Equation~(\ref{viscrs}) can be 
rewritten in the conventional form:
\begin{equation}
\label{viscrs1}
\sigma_{vis}(s) = \sigma(s)B\epsilon(\sqrt{s})(1+\delta(s)),
\end{equation}
where the detection efficiency $\epsilon(\sqrt{s})$ and the radiative 
correction $\delta(s)$  are defined as follows:
\begin{equation}
\epsilon(\sqrt{s}) \equiv \epsilon_r(\sqrt{s},0),
\end{equation}
\begin{equation}
\delta(s) = \frac{\int\limits_{0}^{x_{max}}
\epsilon_r(\sqrt{s},\frac{x\sqrt{s}}{2}) F(s,x,) \sigma(s(1-x))dx}
{\epsilon_r(\sqrt{s},0)\sigma(s)} - 1 .
\end{equation}

Technically the experimental Born cross section is determined as
follows. Using Eq.~(\ref{viscrs}), the energy 
dependence of the measured visible cross section $\sigma_{vis,i} = 
(N_i-N_{bkg,i})/L_i$ is fitted by a theoretical model that describes 
data reasonably well. Here $N_i$, $N_{bkg,i}$, and $L_i$ are
respectively the number of selected data events, the number of background 
events, and  the integrated luminosity for the $i$-th energy interval. 
The fitted parameters of the theoretical model are used to 
calculate the radiative corrections. Then the experimental Born cross
section $\sigma_i$ is calculated using Eq.(\ref{viscrs1}). 

The energy dependence of the Born cross section for the process 
$e^+e^-\to\omega\eta\pi^0$ is parametrized according to the vector meson 
dominance model~\cite{achasov} assuming the $\omega a_0(980)$ 
intermediate state mechanism. The cross section is described by the 
contribution of only one resonance with the mass $m_V$ and width $\Gamma_V$:
\begin{equation}
\label{parcrs}
\sigma(s) = \sigma_V\frac{m^3_V q(s)}{s^{3/2}q(m_V^2)}
\frac{m_V^2 \Gamma_V^2}{(m_V^2-s)^2+s\Gamma_V^2},
\end{equation}
where  $\sigma_V$ is the cross section at $s=m_V^2$ and the 
function $q(s)$ describes the energy dependence of the phase space volume 
of the final state. Far away from the threshold of the reaction 
$e^+e^-\to\omega a_0(980)$, when we can neglect finite widths of the $\omega$
and $a_0(980)$ resonances, $q(s)$ coincides with the $a_0(980)$ momentum.

\begin{figure}
\includegraphics[width=0.45\textwidth]{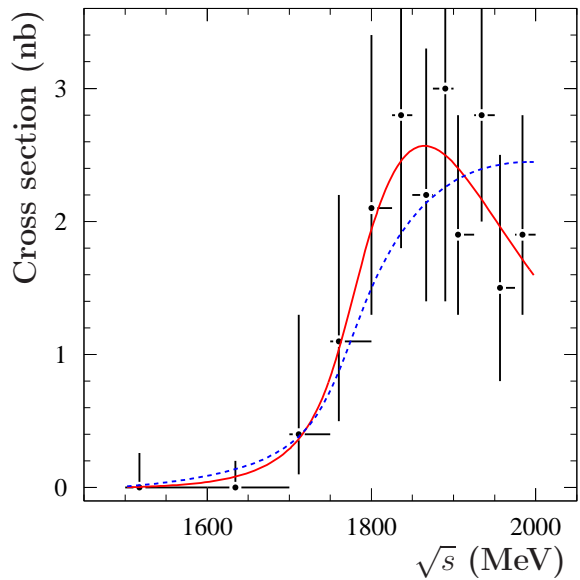}
\caption{ \label{crs1}
The cross section for the process $e^+e^-\to\omega\eta\pi^0$ measured in 
this work. The solid (dashed) curve shows the result of the fit with (without)
a resonance contribution.
}
\end{figure}

Free fit parameters are $\sigma_V$, $M_V$, and 
$\Gamma_V$. The resulting curve is shown in Fig.~\ref{crs1} 
along with the values of the Born cross section
calculated according to Eq.(\ref{viscrs1}).
The obtained values of the mass and width of the resonance, 
$M_V = 1815^{+44}_{-118}$~MeV and $\Gamma_V = 349^{+393}_{-118}$~MeV,
are statistically consistent with the $\rho(1700)$-resonance 
parameters~\cite{pdg}. We also perform a phase-space fit without a resonance 
contribution [$\Gamma_V\to\infty$ in Eq.~(\ref{parcrs})]. The fit 
shown in Fig.~\ref{crs1} by the dashed curve also describes data well.
The significance of the resonance contribution estimated from the difference
of the logarithmic likelihoods of the fits with and without
resonance is about $1.2\sigma$.

The numerical values of the Born cross section and radiative corrections are 
listed in Table~\ref{tabl1}. The total systematic uncertainty on the cross 
section is 4.2 \%. It includes the systematic uncertainties on the detection 
efficiency(3\%), luminosity measurement (2.2\%), and radiative correction (2\%).
The latter is estimated by varying the fit parameters within their errors.  
\begin{table}
\caption{\label{tabl1}
The energy interval, integrated luminosity ($L$), number of selected 
events ($N$), estimated number of background events ($N_{bkg}$),  
detection efficiency for $e^+e^-\to\omega\eta\pi^0\to 7\gamma$ events
($\epsilon$), radiative correction ($\delta+1$), and 
$e^+e^-\to\omega\eta\pi^0$ Born cross section ($\sigma$). The shown 
cross-section errors are statistical. The systematic error is 4.2\%. The 90\%
confidence level upper limits are listed for the first two energy intervals.}
\begin{ruledtabular}
\begin{tabular}{cccccccc}
$\sqrt{s}$ (MeV) & $L$ (nb$^{-1}$) & $N$ & $N_{bkg}$ & $\epsilon$ (\%) & $\delta+1$ & $\sigma$ (nb) \\ 
\hline
1500 $\div$ 1600  & 5888 &  0  & 0.02 & 2.85 & 0.779 & $< 0.26$    \\
1600 $\div$ 1700  & 5004 &  0  & 0.21 & 4.03 & 0.815 & $< 0.20$    \\
1700 $\div$ 1750  & 2261 &  1  & 0.13 & 4.25 & 0.813 & $0.4^{+0.9}_{-0.3}$ \\
1750 $\div$ 1800  & 2392 &  3  & 0.09 & 4.29 & 0.810 & $1.1^{+1.1}_{-0.6}$ \\
1800 $\div$ 1825  & 2373 &  6  & 0.09 & 4.61 & 0.823 & $2.1^{+1.3}_{-0.8}$ \\
1825 $\div$ 1850  & 1897 &  7  & 0.05 & 5.02 & 0.841 & $2.8^{+1.5}_{-1.0}$ \\
1850 $\div$ 1875  & 2527 &  8  & 0.06 & 5.34 & 0.855 & $2.2^{+1.1}_{-0.8}$ \\
1875 $\div$ 1900  &  662 &  3  & 0.02 & 5.58 & 0.865 & $3.0^{+2.9}_{-1.6}$ \\
1900 $\div$ 1925  & 3459 &  9  & 0.07 & 5.12 & 0.871 & $1.9^{+0.9}_{-0.6}$ \\
1925 $\div$ 1950  & 2361 & 11  & 0.05 & 5.98 & 0.880 & $2.8^{+1.1}_{-0.8}$ \\
1950 $\div$ 1975  & 2077 &  5  & 0.06 & 5.64 & 0.887 & $1.5^{+1.0}_{-0.7}$ \\
1975 $\div$ 2000  & 2682 &  9  & 0.06 & 6.14 & 0.893 & $1.9^{+0.9}_{-0.6}$ \\ \hline
\end{tabular}
\end{ruledtabular}
\end{table}

\section{Conclusion}
We have analyzed data collected in the experiment with the SND detector at 
the $e^+e^-$  collider VEPP-2000 in the c.m. energy 
range from 1.05 to 2.00~GeV. In the seven-photon final state, events of the
process $e^+e^- \to \eta\pi^0\pi^0\gamma$ have been separated. Most of these 
events come from the process $e^+e^- \to \omega\eta\pi^0$.
We have measured the cross section for this process for the first time. 
It has a threshold at 1.75 GeV.  The cross section value in the energy range
1.8-2.0 GeV is about 2~nb, approximately 5\% of the total hadronic 
cross section in this energy range. From the analysis of the $\eta\pi^0$ 
invariant mass spectrum, it has been found that the dominant mechanism of the 
reaction $e^+e^- \to \eta\pi^0\pi^0\gamma$ is the $\omega a_0(980)$ 
intermediate state. 

\section{ACKNOWLEDGMENTS}
Part of this work related to the photon reconstruction algorithm in the
electromagnetic calorimeter for multiphoton events is supported
by the Russian Science Foundation (project No. 14-50-00080).

\end{document}